\documentclass[aps,twocolumn,footinbib,prb]{revtex4-1}
\usepackage[colorlinks=true,linkcolor=blue,citecolor=blue]{hyperref}
\usepackage{graphicx}
\usepackage{amsmath}
\usepackage{amssymb}
\begin{document}
\title{Collective transport in the discrete Frenkel-Kontorova model}
\author{T D Swinburne}
\email{tds110@ic.ac.uk}
\affiliation{Department of Physics, Imperial College London, Exhibition Road, London SW7 2AZ, UK}
\affiliation{EURATOM/CCFE Fusion Association, Culham Centre for Fusion Energy, Abingdon, Oxfordshire OX14 3DB, UK}
\date{\today}
\begin{abstract}
Through multiscale analysis of the adjoint Fokker-Planck equation, strict bounds are derived for the center of mass diffusivity of an overdamped harmonic chain in a periodic potential, often known as the discrete Frenkel-Kontorova model. Significantly, it is shown that the free energy barrier is a lower bound to the true finite temperature migration barrier for this general and popular system. Numerical simulation confirms the analysis, whilst effective migration potentials implied by the bounds are employed to give a surprisingly accurate prediction of the non-linear response.
\end{abstract}
\maketitle
\begin{figure}
\frame{\includegraphics[width=0.45\textwidth]{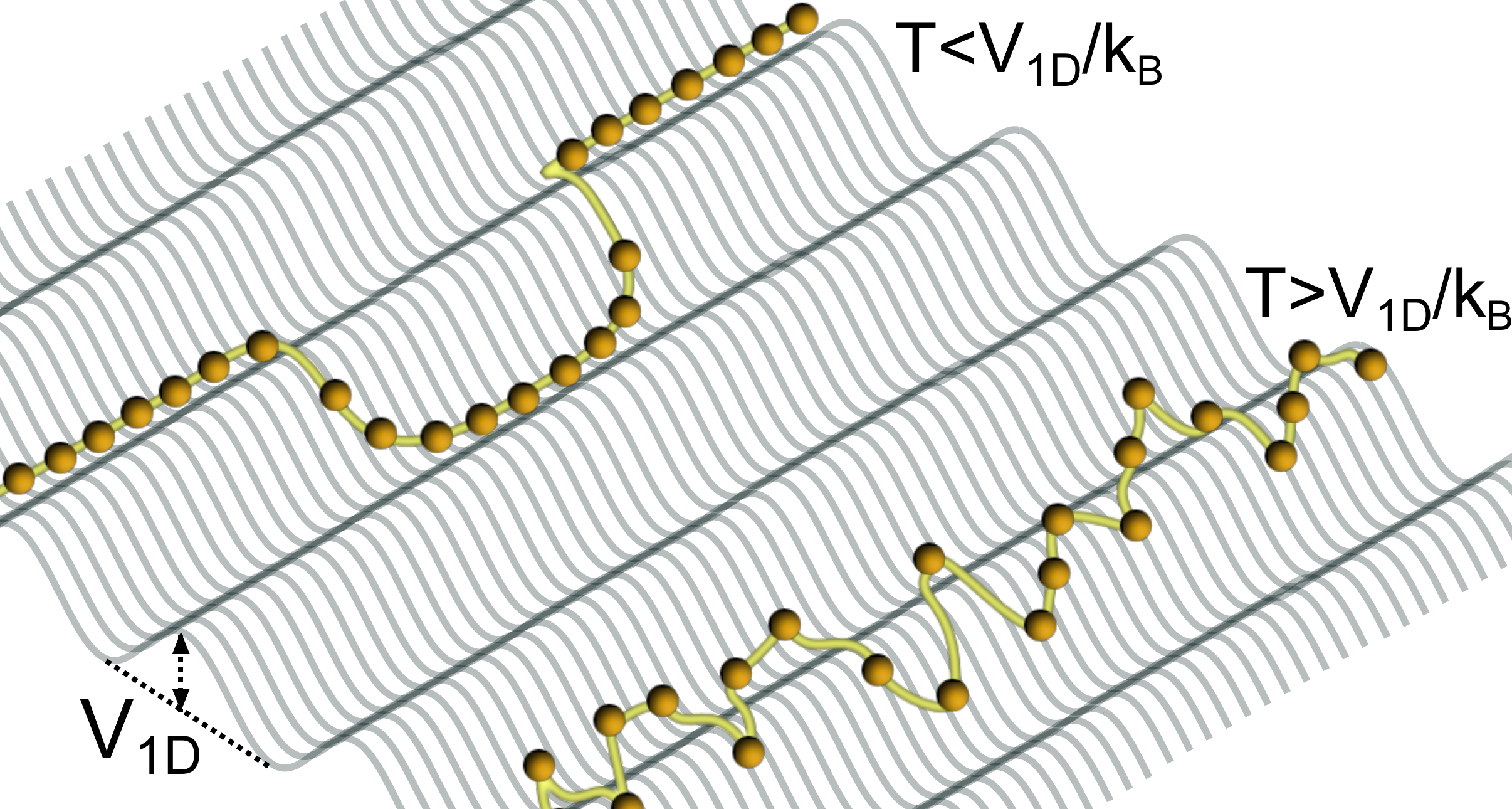}}
\caption{(Color online) A Frenkel-Kontorova chain. At low temperature (left) the chain moves through the kink mechanism, whilst at high temperature (right) internal fluctuations destroy any migration barrier.\label{FKC}} 
\end{figure}

A chain of harmonically coupled particles, each executing one dimensional stochastic motion in a periodic potential, is one of the most extensively studied  examples of many-body, non-linear dynamics. First studied by Prandtl\cite{prandtl1928} and Dehlinger\cite{dehlinger1929} though often named after later work by Frenkel and Kontorova\cite{kontorova1938}, the rich, kink bearing phenomenology has found application in dislocation theory\cite{swinburne2013,Proville2012}, polymer dynamics\cite{Wall2005}, molecular combustion\cite{derenyi1995}, Josephson junctions\cite{Fedorov2009}, spin chains\cite{fesser1980}, earthquakes\cite{gershenzon2009} and many other areas for decades\cite{Braun2004,hanggi2009}. In the general case, illustrated in Figure \ref{FKC}, a Frenkel-Kontorova (FK) chain of $N$ particles with one dimensional positions ${\bf x} = \{x_n\}_{n=1}^N$ has a potential energy
\begin{equation}
U({\bf x}) = \frac{1}{2}{\bf x}\cdot{\bf K}\cdot{\bf x} + V({\bf x}),\label{pot}
\end{equation}
where $\bf K$ is a positive semi definite matrix representing the harmonic interaction and $V(\bf x)$ is simply a sum of one dimensional periodic potentials $V_{\rm 1D}(x) = V_{\rm 1D}(x+L)$
\begin{equation}
V({\bf x}) = \sum_{n=1}^N V_{\rm 1D}(x_n).
\end{equation}The system is completed with chain boundary conditions, which will be periodic in the following. As the FK chain traditionally models the collective motion of some generalized charges, it is of central interest to know the transport properties of the chain center of mass 
\begin{equation}
\bar{x}={\sum}_n x_n/N,
\end{equation}
in particular the diffusivity $D$ and by Einstein's relation the linear response mobility $\beta D$, where $\beta=1/k_{\rm B}T$. Whilst it is known\cite{SCALING} that the center of mass is diffusive at asymptotic time, the actual value of the diffusion constant $D$ has only been approximately evaluated for some special cases, in particular for long, continuous lines at low temperature, where the system has been considered as a dilute kink gas\cite{Alexander1993,Habib2000}. In contrast, many applications of interest are to highly discrete chains over a wide temperature range which are often short due to either physical\cite{Fedorov2009,Dudarev2011} or computational\cite{Proville2012,Gilbert} restrictions. In this paper I derive rigorous upper and lower bounds for $D$, giving important context for existing approaches such as transition state theory\cite{Kramers} and providing rigorous results for many body diffusive transport.\\
Through comparing the bounds to the well known point particle result\cite{Lifson,Risken} it is shown that the upper bound represents diffusion in the free energy landscape of $\bar x$. The free energy barrier is often used as the finite temperate migration barrier\cite{kumar1992}; these results show that this will always give an overestimate for the transport properties of the FK chain, an important result given the generality of this widely applied model.\\ 
The paper is structured as follows. In section \ref{sec:afp} the adjoint Fokker-Planck equation\cite{Zwanzig} is recalled, then multiscale analysis is employed to perform a diffusive rescaling in section \ref{sec:hom}, deriving an one dimensional evolution equation for the center of mass. The Cauchy-Schwartz inequality is then used to derive strict upper and lower bounds for the effective diffusion constant $D$. In section \ref{sec:KG} I investigate limiting cases of the exact bounds, present numerical results in section \ref{sec:NM} and propose a non-linear response through analogy to the famous point particle result of Stratonovich\cite{kusnev,stratonovich1967} in section \ref{sec:nlr}, where surprisingly accurate results are found.\\
\section{Adjoint Fokker-Planck equation}\label{sec:afp}
For later manipulations it will be beneficial to transform to a coordinate system which distinguishes the center of mass. This is acheived by diagonalising the interaction matrix $\bf K$, which will always have non-negative eigenvalues $\{\lambda_k\}_{k=1}^{N}$ and an orthonormal eigenbasis $\{\hat{\bf v}_k\}_{k=1}^{N}$. By the requirement that the interaction energy is unchanged under a rigid translation, there will always be a zero eigenvalue, $\lambda_1=0$, with the corresponding eigenvector $\hat{\bf v}^1$ having every element equal, projecting out the center of mass $\bar x$. The chain configuration vector $\bf x$ becomes
\begin{equation}
{\bf x} = {\bar x}\sqrt{N}\hat{\bf v}_1 +\sum_{k=2}^{N} a_k\hat{\bf v}_k\quad,\quad a_k = {\bf x}\cdot\hat{\bf v}_k,\label{coexp}
\end{equation}
which defines the desired co-ordinate system $({\bar x},\{a_k\})$. The potential energy (\ref{pot}) now reads
\begin{equation}
U({\bar x},\{a_k\}) = \sum_{k=2}^{N}\frac{1}{2}\lambda_ka^2_k + V({\bar x},\{a_k\}),\label{pot2} 
\end{equation}
where the substrate potential is explicitly
\begin{equation}
V({\bar x},\{a_k\}) = \sum_{n=1}^NV_{\rm 1D}\left({\bar x}+\sum_{k=2}^N a_k(\hat{\rm v}_k)_n\right),
\end{equation}
which is clearly periodic in $\bar x$. One may now write down the adjoint Fokker-Planck equation\cite{Zwanzig}, which governs the expected time evolution of a smooth function $\Phi(t;{\bar x},\{a_k\})$ from some initial values $({\bar x},\{a_k\})$. For the investigation of transport properties, the adjoint Fokker-Planck equation is preferable to the Fokker-Planck equation as it is concerned with observables rather than probability densities, but any results may be rigorously transferred between the two presentations, in close analogy to the Schr\"odinger and Heisenberg representations of quantum mechanical operators\cite{Zwanzig}. For the system (\ref{pot2}) the adjoint Fokker-Planck equation reads
\begin{align}
N\beta\gamma\frac{\partial\Phi}{\partial t} = \hat{\rm L}_{\rm aFP}\Phi &\equiv -\beta\frac{\partial U}{\partial \bar{x}}\frac{\partial\Phi}{\partial \bar{x}}+\frac{\partial^2\Phi}{\partial \bar{x}^2}\label{afp}\\
&\,+ N\sum_{k=2}^N\left(-\beta\frac{\partial U}{\partial a_k}\frac{\partial\Phi}{\partial a_k}+\frac{\partial^2\Phi}{\partial a^2_k}\right),\nonumber
\end{align}
where $\hat{\rm L}_{\rm aFP}$ is the adjoint Fokker-Planck operator, $U$ is given by (\ref{pot2}) and $\gamma$ is the friction parameter, which measures the rate of momentum transfer to the heat bath (a factor of $N\beta\gamma$ has been taken to the left hand side of (\ref{afp}) to simplify later notation). For the overdamped limit to be valid, which amounts to a `Born-Oppenheimer' decoupling of position and momentum, $\gamma$ is required to be much greater than the curvatures of $U$\cite{reif}. 
Familiar statistical mechanics arises upon averaging over the initial conditions and asking for the steady state; the condition for the probability density of states $\rho_\infty({\bar x},\{a_k\})$ is
\begin{equation}
0 = \int_{{\bar x},\{a_k\}}  \left({\hat{\rm L}_{\rm aFP}}\Phi\right)\rho_\infty=
\int_{{\bar x},\{a_k\}}\left({\hat{\rm L}^*_{\rm aFP}}\rho_\infty\right)\Phi,\label{invar}
\end{equation}
where $\hat{\rm L}^*_{\rm aFP}$ is the $L^2$ adjoint of $\hat{\rm L}_{\rm aFP}$\cite{ethier2009}, producing the overdamped Fokker-Planck (or Smolchowski) equation.   As is well known, the unique solution is Gibbs' distribution
\begin{equation}
{\hat{\rm L}^*_{\rm aFP}}\rho_\infty = 0\quad\Rightarrow\quad\rho_\infty = e^{-\beta U}/Z,\label{gibbs}
\end{equation}
where $Z$ is the partition function. Due to the periodicity of $U$ in $\bar x$, the Fokker-Planck operator and thus any unique solution will also be periodic in $\bar x$; however, for the steady state (\ref{gibbs}) to exist in this case we require ${\bar x}\in[0,L]$, which clearly forbids diffusion. To extract a diffusion constant we will use multiscale analysis in the next section to investigate the diffusive dynamics of a coarse grained center of mass $\bar\chi\in[-\infty,\infty]$, which is asymptotically independent of $\bar x \in [0,L]$ as the scale separation diverges.\\ 

Throughout this paper integrals over ${\bar x}$ and the $\{a_k\}$ will be denoted as $\int_{{\bar x},\{a_k\}}$, with the bounds of integration being $[0,L]$ for $\bar x$ and $[-\infty,\infty]$ for each $a_k$. Integrals over only the $\{a_k\}$ will be denoted as $\int_{\{a_k\}}$, again integrating over $[-\infty,\infty]$ for each $a_k$. The proof\cite{ethier2009,Pavliotis2008diffusive,ottobre2011} of ergodicity and the existence of an unique steady state (\ref{gibbs}) for potentials of the form (\ref{pot2}) follows from the quadratic confinement of $\sum_p \lambda_p a^2_p/2$ and the boundedness of $V({\bar x},\{a_p\})$.
\section{Multiscale analysis}\label{sec:hom}
The techniques used in the following are detailed in the recent book by Pavliotis and Stuart\cite{Pavliotis2008}, an accessible introduction which contains extensive references, though it is believed that the present application to a many body system is new material.\\ The central idea behind multiscale analysis is that at long times unbound variables can have unbound expectation values, which will be much larger than any length scale imposed by the potential environment. In the present case the unbound variable is the center of mass $\bar x$, whose variance at asymptotic time diverges linearly and therefore will be much greater than the potential period $L$. As a result, to extract an effective diffusion constant one may work on a coarse grained time and length scale which will be insensitive to details of the underlying potential. This is often what occurs in simulation or experiment; it is acheived analytically through first rescaling time as
\begin{equation}
t\rightarrow \frac{t}{\epsilon^2}\quad,\quad 0<\epsilon\ll1,\label{rescale}
\end{equation}
then identifying the `slow' spatial variable
\begin{equation}
{\bar \chi}=\epsilon{\bar x}\label{rescale2}.
\end{equation}
Such an approach was first used by Hilbert to investigate hydrodynamic limits of the Boltzmann equation\cite{Hilbert}. On a coarse time scale, of order one as $\epsilon\to0$, the dynamics of $\bar x$ and the $\{a_k\}$ will be massively faster than those of $\bar \chi$. In particular, as $\bar x$ moves in a periodic potential it will fluctuate extremely rapidly, so that as $\epsilon\to0$, $\bar \chi$ and $\bar x$ are scale separated and become independent variables. By this definition, the potential $U$ only depends on $({\bar x},\{a_k\})$ as the fast variables will only have a homgenised affect on the slow variable $\bar \chi$. Employing the transformations (\ref{rescale}), (\ref{rescale2}) and using the chain rule, consider functions $\Phi^\epsilon({\bar \chi},{\bar x},\{a_k\})$ which solve the adjoint Fokker-Planck equation\cite{Pavliotis2008}
\begin{equation}
N\beta\gamma\frac{\partial\Phi^\epsilon}{\partial t} =
\frac{\partial^2\Phi^\epsilon}{\partial \bar{\chi}^2}
+\frac{2}{\epsilon}\frac{\partial^2\Phi^\epsilon}{\partial{\bar x}\partial {\bar\chi}}
-\frac{\beta}{\epsilon}\frac{\partial U}{\partial \bar{x}}\frac{\partial\Phi^\epsilon}{\partial \bar{\chi}}
+\frac{1}{\epsilon^2}\hat{\rm L}_{\rm aFP}\Phi^\epsilon,\label{mafp}
\end{equation}
where $\hat{\rm L}_{\rm aFP}$ is definied in equation (\ref{afp}) and acts only on $({\bar x},\{a_k\})$. In the absence of any potential landscape, equation (\ref{mafp}) would represent free diffusion for $\bar \chi$, justifying the scaling operations (\ref{rescale}) and (\ref{rescale2}). By the aforementioned periodicity of $U$, $\Phi^\epsilon$ will be periodic in $\bar x$\cite{PERnote} meaning $\bar x$ can be constrained to take values in the interval $[0,L]$. To look for an explicit solution, perform a multiscale expansion of $\Phi^\epsilon$ in orders of the small parameter $\epsilon$,
\begin{equation}
\Phi^\epsilon = \Phi_0+\epsilon\Phi_1+\epsilon^2\Phi_2 + ...,\label{msexp}
\end{equation}
where at asymptotic time the solution will be given by $\Phi_0$. Substituting (\ref{msexp}) into (\ref{mafp}) produces a hierarchy of equations in orders of $(1/\epsilon)$, reading
\begin{align}
O\left(\frac{1}{\epsilon^2}\right)&: {\hat{\rm L}}_{\rm aFP}\Phi_0 =  0,\label{o2}\\
O\left(\frac{1}{\epsilon}\right)&:
{\hat {\rm L}}_{\rm aFP}\Phi_1 - \beta\frac{\partial U}{\partial {\bar x}}\frac{\partial \Phi_0}{\partial \bar{\chi}} +2\frac{\partial^2 \Phi_0}{\partial {\bar x}\partial{\bar\chi}}= 0,\label{o1}\\
O\left({1}\right)&:
{\hat{\rm L}}_{\rm aFP}\Phi_2 + \frac{\partial^2 \Phi_0}{\partial \bar{\chi}^2}+2\frac{\partial^2 \Phi_1}{\partial \bar{\chi}\partial {\bar x}}\nonumber\\
&\quad\quad\quad-\beta\frac{\partial U}{\partial {\bar x}} \frac{\partial \Phi_1}{ \partial \bar{\chi}} = N\beta\gamma\frac{\partial\Phi_0}{\partial t}\label{o0}.
\end{align}
To reduce these hierarchy of equations into a single effective equation for $\Phi_0$ it is required to solve Poisson equations of the form
\begin{equation}
\hat{\rm L}_{\rm aFP}f({\bar\chi},\bar{x},\{a_k\},t)=g({\bar\chi},\bar{x},\{a_k\},t),\label{operex}
\end{equation}
for two smooth functions $f$ and $g$ which satify the normalisation condition
\begin{equation}
\int_{{\bar x},\{a_k\}}\rho_\infty({\bar x},\{a_k\})|f({\bar\chi},\bar{x},\{a_k\},t)|^2<\infty\label{opernorm},
\end{equation}
where $\rho_\infty$ is given by (\ref{gibbs}), and is a restatement of the requirement that the expectation values are finite after a finite time.  Due to the smoothness of the parabolic operator $\hat{\rm L}_{\rm aFP}$, it is well known\cite{ethier2009,Pavliotis2008,hairer2008ballistic}  that (\ref{operex}) has a unique solution (up to constants) if and only if
\begin{equation}
\int_{{\bar x},\{a_k\}}\rho_\infty({\bar x},\{a_k\}){g({\bar\chi},\bar{x},\{a_k\},t)} = 0\label{opernorm2}.
\end{equation}
This condition may be justified by considering acting on (\ref{operex}) with $\rho_\infty$ and integrating over the support of the exponent, which as defined above is $[0,L]$ for $\bar x$ and  $(-\infty,\infty)$ for $\{a_{k}\}$. Providing the normalisation condition holds, use (\ref{invar}) and (\ref{operex}) to show
\begin{equation}
\int_{{\bar x},\{a_k\}}\rho_\infty g=\int_{{\bar x},\{a_k\}}f\hat{L}^*_{\rm aFP}\rho_\infty=0.
\label{opernorm3}
\end{equation}
Now apply the conditions (\ref{opernorm}), (\ref{opernorm2}) to the equations (\ref{o2}), (\ref{o1}), (\ref{o0}), which are all of the form (\ref{operex}). The first equation, (\ref{o2}), acts on $({\bar x},\{a_k\})$ and thus by uniqueness $\Phi_0$ is a function only of $\bar\chi$ and $t$,
\begin{equation}
\Phi_0({\bar\chi},\bar{x},\{a_k\},t) = \Phi_0(\bar{\chi},t).
\end{equation}
Condition (\ref{opernorm}) requires that for a solution of (\ref{o1}) to exist
\begin{equation}
-\beta\int_{{\bar x},\{a_k\}} \rho_\infty\frac{\partial U}{ \partial {\bar x}}\frac{\partial \Phi_0}{\partial \bar{\chi}}
=
\left(\int_{{\bar x},\{a_k\}} \frac{\partial\rho_\infty}{\partial {\bar x}}\right)\frac{\partial \Phi_0}{\partial \bar{\chi}} = 0,
\end{equation}
which is clearly satisfied as $\rho_\infty$ is periodic in $\bar x$. This allows one to try a separated variable solution of the form
\begin{equation}
\Phi_1(\bar{\chi},{\bar x},\{a_k\},t) = \phi({\bar x},\{a_k\})\frac{\partial \Phi_0}{\partial \bar{\chi}},
\end{equation}
which when substituted into (\ref{o1}) gives
\begin{equation}
\hat{\rm L}_{\rm aFP}\phi = \frac{\partial U}{\partial {\bar x}}\label{hPoiss}.
\end{equation}
Finally, apply the condition (\ref{opernorm}) to (\ref{o0}). Multiply (\ref{o0}) by $\rho_\infty$ and integrate over all $({\bar x},\{a_k\})$. The $\Phi_2$ term disappears by (\ref{opernorm3}), to that after an integration by parts,
\begin{equation}
\frac{\partial\Phi_0}{\partial t} = \left(\int_{{\bar x},\{a_k\}}\rho_\infty \left(1+\frac{\partial\phi}{\partial {\bar x}}\right)\right)\frac{\partial^2 \Phi_0}{\partial\bar{\chi}^2}.\label{DE}
\end{equation}
Equation (\ref{DE}) is easily recognisable as an (adjoint) free diffusion equation in $\bar{\chi}$ with an effective diffusion constant
\begin{equation}
D = \frac{1}{N\beta\gamma}\int_{q,\{a_{p>0}\}}\rho_\infty \left(1+\frac{\partial\phi}{\partial q}\right).\label{D2}
\end{equation}
It simple to show that with $\Phi_0$=$\left<{\bar \chi}^2\right>$ one obtains $\left<{\bar \chi}^2\right>$=$2Dt$. To simplify the following presentation, I work with the reduced diffusivity $\tilde{D}=N\beta\gamma{D}$. Using (\ref{hPoiss}) and (\ref{afp}), $\tilde D$ may be written 
\begin{equation}
\tilde{D} = \int_{{\bar x},\{a_k\}}\rho_\infty \left(\left(1+\frac{\partial\phi}{\partial \bar x}\right)^2+
\sum_{k=2}^N\left(\frac{\partial\phi}{\partial a_k}\right)^2\right).\label{D3}
\end{equation}
I shall use both expressions (\ref{D2}), (\ref{D3}) in the following section where the Cauchy-Schwartz inequality\cite{Abrams} (CSI) is employed to obtain upper and lower bounds for $\tilde{D}$. Using the normalisation condition (\ref{opernorm}), the CSI reads
\begin{equation}
\left(\int_{{\bar x},\{a_k\}}\ \rho_\infty fg\right)^2\leq \left(\int_{{\bar x},\{a_k\}}\ \rho_\infty f^2\right)\left(\int_{{\bar x},\{a_k\}}\rho_\infty g^2\right).\label{CS1}
\end{equation} 
For the special case here, where the functions under consideration are smooth, periodic and bounded in $\bar x$, one may again use (\ref{opernorm}) to write (See Appendix \ref{sec:app})
\begin{equation}
\left(\int_{\{a_k\}}\ \rho_\infty fg\right)\leq \left(\int_{\{a_k\}}\ \rho_\infty f^2\right)\left(\int_{\{a_k\}}\ \rho_\infty g^2\right),\label{CS2}
\end{equation}
which holds for all $\bar x\in[0,L]$. To proceed, note that for any real function $\phi$ the following inequality is always satisfied
\begin{align}
\tilde{D} &=\int_{{\bar x},\{a_k\}}\rho_\infty \left(\left(1+\frac{\partial\phi}{\partial \bar x}\right)^2+
\sum_{k=2}^N\left(\frac{\partial\phi}{\partial a_k}\right)^2\right)\nonumber\\
&\geq \int_{{\bar x},\{a_k\}}\rho_\infty \left(1+\frac{\partial\phi}{\partial \bar x}\right)^2.\label{ineq0}
\end{align}
Also define the `harmonic chain' partition function
\begin{equation}
Z_\lambda=\int_{\{a_k\}}e^{-\beta\sum_k\lambda_ka^2_k/2}=\prod_{k=2}^N\sqrt{\frac{\pi}{\beta\lambda_k}},
\end{equation}
allowing one to write a useful quantity, a conditional average of $\exp(\pm\beta V)$ over all configurations with a center of mass $\bar x$ as
\begin{equation} 
\langle e^{\pm\beta V};\bar{x}\rangle = Z^{-1}_\lambda\int_{\{a_k\}} e^{\pm\beta V({\bar x},\{a_k\})-\beta\sum_k\lambda_ka^2_k/2},\label{haexp}
\end{equation}
meaning in particular that
\begin{equation}
\oint_{\bar x}\langle e^{-\beta V};\bar{x}\rangle = Z^{-1}_\lambda\int_{{\bar x},\{a_k\}} e^{-\beta U({\bar x},\{a_k\})} = \frac{Z}{Z_\lambda},
\end{equation}
where $U({\bar x},\{a_k\})$ is given by (\ref{pot2}) and $Z$ is the full partition function. To obtain a lower bound for $\tilde{D}$, use the fact that $\rho_\infty \exp(\beta V)$ is independent of $\bar x$ and the periodicity of $\phi$ in $\bar x$ to give
\begin{equation}
\int_{{\bar x},\{a_k\}}\left(1+\frac{\partial\phi}{\partial \bar x}\right)\rho_\infty e^{\beta V} = L\frac{Z_\lambda}{Z}.\label{iden1}
\end{equation}
Applying the Cauchy-Schwartz inequality (\ref{CS1}) to (\ref{iden1}), using (\ref{ineq0}), produces the first main result, a strict lower bound for the center of mass diffusivity,
\begin{equation}
D \geq {D}_{\rm L} = \frac{L^2/N\beta\gamma}{\oint_{\bar x}\langle e^{-\beta V};\bar{x}\rangle{\rm d}{\bar x}\oint_{\bar y}\langle e^{\beta V};\bar{y}\rangle{\rm d}{\bar y}}.\label{lowbound}
\end{equation}
To derive an upper bound for $D$, multiply (\ref{hPoiss}) by $\rho_\infty$ and integrate over all $\{a_k\}$, but crucially not $\bar x$, to obtain
\begin{equation}
\int_{\{a_k\}}\left(1+\frac{\partial\phi}{\partial \bar x}\right)\rho_\infty = \frac{\tilde{D}}{L},\label{nhb}
\end{equation}
where I have integrated by parts and used (\ref{D2}). Applying the second Cauchy-Schwartz inequality (\ref{CS1}) to (\ref{nhb}) and using (\ref{ineq0}) results in
\begin{equation}
\frac{\tilde{D}^2}{L^2} \leq \frac{\tilde{D}}{L}\frac{
\langle e^{-\beta V};\bar{x}\rangle}{\oint_{\bar y}\langle e^{-\beta V};\bar{y}\rangle{\rm d}{\bar y}}.
\end{equation}
Whilst integration over $\bar x$ simply shows that the reduced diffusivity ${\tilde D}\leq1$, dividing both sides by $\langle \exp({-\beta V});\bar{x}\rangle$ then integrating produces the second main result, a strict upper bound for the center or mass diffusivity,
\begin{equation}
D\leq D_{\rm U} = \frac{L^2/ N\beta\gamma}{\oint_{\bar x}\langle e^{-\beta V};\bar{x}\rangle^{-1}{\rm d}{\bar x}\oint_{\bar x}\langle e^{-\beta V};\bar{y}\rangle{\rm d}{\bar y}}.\label{highbound}
\end{equation}
Both bounds benefit from a comparison to the well known diffusivity of a point particle moving in an one dimensional periodic potential $V_{\rm 1D}(x)=V_{\rm 1D}(x+L)$\cite{Lifson,Risken}
\begin{equation}
D_{\rm 1D} = \frac{L^2/\gamma\beta}{\int_0^L e^{-\beta V_{\rm 1D}(x)}{\rm d}x\int_0^L e^{+\beta V_{\rm 1D}(y)}{\rm d}y}.\label{LJ}
\end{equation}
Using (\ref{LJ}) and the bounds (\ref{lowbound}), (\ref{highbound}) it is simple to show\cite{POTnote} that to within unimportantant constants, the lower ($\rm L$) and upper ($\rm U$) bounds are equivalent to the diffusivity of a point particle moving in the periodic potential
\begin{equation}
F_{\rm L,U}(\bar x) = \pm k_{\rm B}T\ln\langle e^{\pm\beta V};{\bar x}\rangle.\label{Epot}
\end{equation}
In particular, from the definition (\ref{haexp}), $\langle\exp(-\beta V);{\bar x}\rangle$ may be written as $Z^{-1}_\lambda\int_{\{a_k\}}\exp(-\beta U)$, so that $F_{\rm U}$ is the Helmholtz free energy landscape of the center of mass\cite{landau1975classical}. As one may extract the free energy from simulation through a simple histogram method\cite{kumar1992} it has become a popular measure of a finite temperature migration barrier, so it is significant that these results show $F_{\rm U}$  to be a lower bound to the true energy barrier experienced by this many body system. I now investigate limiting cases and present simulation results to validate the above analysis.

\section{Limiting Cases}\label{sec:KG}
\begin{figure}
\includegraphics[width=0.48\textwidth]{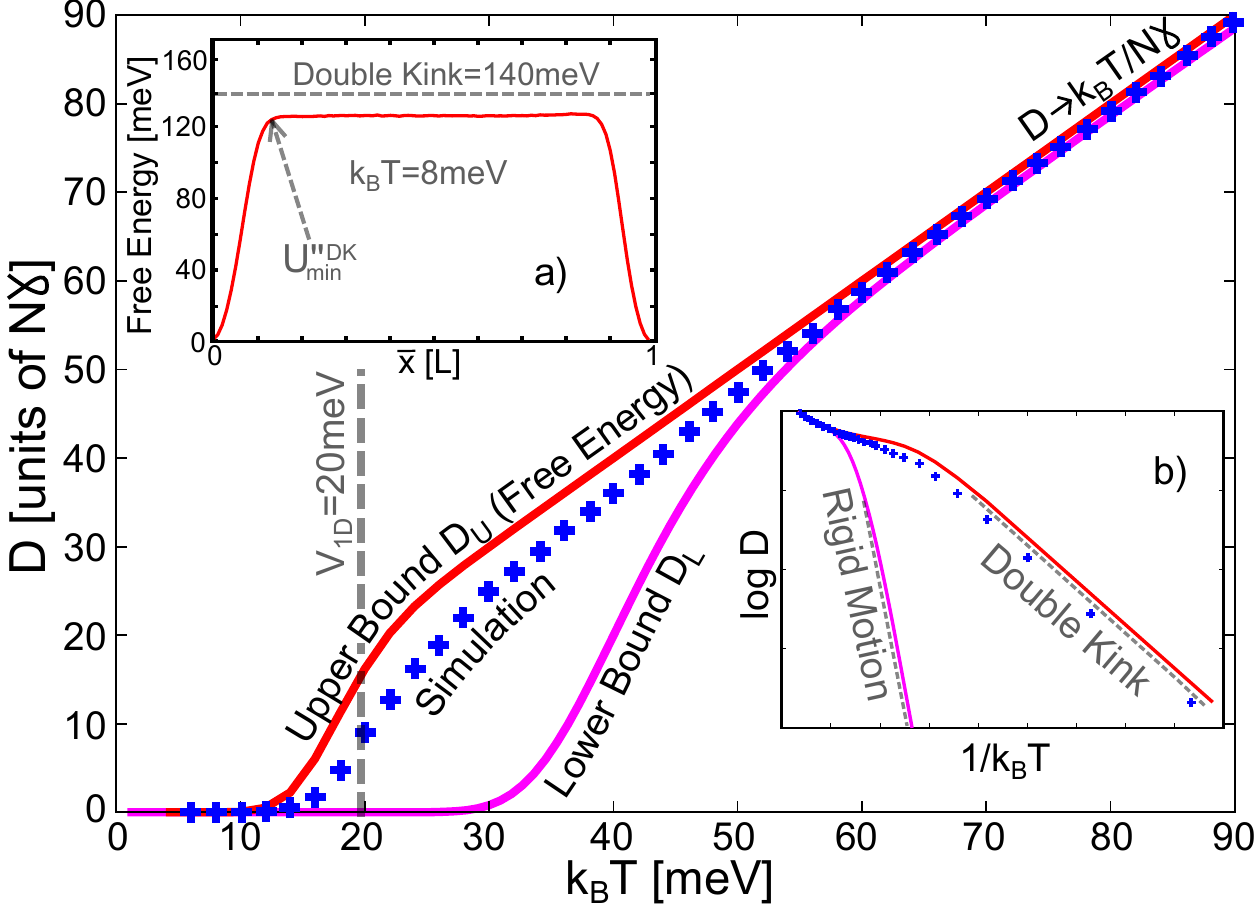}
\caption{(Color online) Diffusivity of a 40 particle Sine-Gordon chain. The upper and lower bounds, equations (\ref{highbound}) and (\ref{lowbound}), agree with simulation and (\ref{hightemp}) at high temperature and capture many important features at intermediate temperature. The diffusivity rises sharply once the thermal energy is greater than the particle barrier $|V_{\rm 1D}|$ (see main text). Inset a): The free energy barrier at low temperature. After a sharp nucleation period, the plateau represents kink pair separation. When the kink energy is comparable to the particle barrier, the plateau energy oscilates with the kink migration barrier\cite{swinburne2013,KMnote}. Inset b): Arrhenius plot of the diffusivity along with the low temperature limits (\ref{hbsd}) and (\ref{lbsd}). The upper bound gives the correct kink pair activation energy.\label{sim_plot}}
\end{figure}

In the low temperature limit $\beta \to \infty$, one may evalutate the integrals over $\{a_k\}$ in the definition (\ref{haexp}) of $\langle\exp(\pm\beta V);{\bar x}\rangle$ by the method of steepest descents\cite{debye1909}. These evaluations can then be used in a steepest descents evaluation of the bounds (\ref{lowbound}), (\ref{highbound}).\\ 
As it has been seen that $\langle\exp(-\beta V);{\bar x}\rangle$ may be written as $Z^{-1}_\lambda\int_{\{a_k\}}\exp(-\beta U)$, at each value of $\bar x$ the integrand will be dominated by the set of coordinates $\{a^{\rm min}_k(\bar x)\}$ which minimise $U$, with a set of $N-1$ second derivatives\cite{CURnote} $\{\omega_k({\bar x})\}_{k=2}^{N}$. As a result the conditional average becomes
\begin{equation}
\langle e^{- \beta V};{\bar x}\rangle\xrightarrow[\beta \to \infty]{}\prod_{k=2}^N\sqrt{\frac{\lambda_k}{\omega_k(\bar x)}}e^{-\beta U_{\rm min}(\bar x)},\label{fsde}
\end{equation}
where $ U_{\rm min}(\bar x)$ is the minimum energy of the system at a given value of $\bar x$. For a sufficiently long and stiff chains (where the largest eigenvalue of $\bf K$ is much greater than the magnitude of the on site potential, resulting in a wide, smooth kink profile) this will be the kink anti-kink pair energy $E_{\rm DK}$ for $\bar x={\bar x}^{\rm DK} \gtrsim 2w_k/N$, where $w_k$ is the kink width, unless the structure of $\bf K$ will give a long range kink interaction\cite{Braun2004}. Additionally, one second derivative, say $\omega^{\rm DK}_2$, will become of order $1/N$ due to the vanishingly small kink pair translation barrier\cite{[{An explicit evaluation for the continuum Sine-Gordon limit is given in }] Ohsawa}. At ${\bar x} = 0$ the chain will be straight, with curvatures
\begin{equation}
\omega_k(0) = V^{''}_{\rm 1D}(0)+\lambda_k.
\end{equation}
One may now evaluate the integrals of $\langle\exp(-\beta V);{\bar x}\rangle$ and its inverse in the upper bound (\ref{highbound}), also by steepest descents at low temperature, which will be dominated by the maximum and minimum values of (\ref{fsde}) respectively. Letting the Goldstone mode $\omega^{\rm DK}_2$ vanish as $1/N$ and recognising that $U^{''}_{\rm min}(0)=N V^{''}_{\rm 1D}(0)$, the low temperature upper bound reads
\begin{equation}
D_{\rm U}
\to
\frac{\sqrt{\pi V^{''}_{\rm 1D}(0)|U_{\rm min}^{''\rm DK}|}}{\gamma\sqrt{\beta}}
\frac{\prod_{k=2}^N\sqrt{\lambda_k+V^{''}_{\rm 1D}(0)}}{\prod_{k=3}^N\sqrt{\omega^{\rm DK}_k}}e^{-\beta E_{\rm DK}},\label{hbsd}
\end{equation}
where $|U_{\rm min}^{'' \rm DK}|$ is the largest negative curvature of $U_{\rm min}$ (see inset a) of Figure (\ref{sim_plot})). This expression is exactly the Arrhenius result of Kramers' transition state theory\cite{Kramers,Kramers1940}, with a length $(N)$ independent prefactor. As shown in section \ref{sec:nlr}, when driving the chain with a homogeneous bias $f$ the center of mass feels a force of $Nf$, meaning that the linear response drift velocity $Nf\beta D_{\rm U}$ is proportional to the length $N$, a recognised signature of the kink pair mechanism when the kink migration barrier vanishes\cite{swinburne2013}.\\
The lower bound (\ref{lowbound}) requires a steepest descents evaluation of
\begin{equation}
\langle e^{\beta V};\bar{x}\rangle = Z^{-1}_\lambda\int_{\{a_k\}} e^{\beta V({\bar x},\{a_k\})-\beta\sum_k\lambda_ka^2_k/2},
\end{equation}
at each value $\bar x$, which as $V>0$ is dominated by the straight line $a_k=0$, $k=2,3,..N$ as $\beta\to\infty$. As a result the low temperature limit for $D_{\rm L}$ reads
\begin{align}
D_{\rm L}&
\to
\frac{\sqrt{V^{''}_{\rm 1D}(0)|V^{''}_{\rm 1D}(L/2)|}}{\gamma}
e^{-\beta N|V_{\rm 1D}|}
\nonumber\\
&\times{\prod_{k=2}^N}
\sqrt{\left(1+\frac{V^{''}_{\rm 1D}(0)}{\lambda_k}\right)\left(1+\frac{|V^{''}_{\rm 1D}(L/2)|}{\lambda_k}\right)}
,\label{lbsd}
\end{align}
as appropriate for essentially rigid motion. At high temperature, as $\beta\to0$, one may perform an expansion of $\pm k_{\rm B}T\ln\langle\exp(\pm\beta V);{\bar x}\rangle$ in orders of $\beta|V_{\rm 1D}|$, being a cumulant expansion for the effective potential\cite{Reichl2009,CONVnote}.  The real periodic on-site potential is expanded in a Fourier series
\begin{equation}
V_{\rm 1D}(x) = \rm{Re}\sum_{p\in\mathbb{Z}}\tilde{V}_pe^{i2\pi p x/L}
\end{equation}
and then use identities of Gaussian integrals and the definition (\ref{Epot}) to write, to order $|\beta V_{\rm 1D}|$
\begin{equation}
F_{\rm L,U}(\bar x) \xrightarrow[\beta \to 0]{}
N\rm{Re}\sum_{p\in\mathbb{Z}}\tilde{V}_pe^{i2\pi p x/L}e^{-p^2\sigma},\label{hightemp}
\end{equation}
where $\sigma=\sum_k 4\pi^2k_{\rm B}T/L^2\lambda_k$ is the mean squared fluctuation of a free harmonic chain\cite{Dudarev2011}. As $\sigma$ increases linearly with $T$, both bounds converge to an effective migration potential which attenuates exponentially fast with increasing temperature. The condition\cite{CONVnote} for convergence of this expansion is $\beta|V_{\rm 1D}|<1$
which can occur at temperatures well below the kink pair energy $\sim\sqrt{\lambda_{\rm max}|V_{\rm 1D}|}$, where $\lambda_{\rm max}$ is the largest eigenvalue of $\bf K$\cite{Braun1998}.\\
\section{Stochastic Simulation}\label{sec:NM}
To test these limiting expressions, consider the Sine-Gordon chain, a special case of (\ref{pot}),
\begin{equation}
U({\bf x}) = \sum_{i=1}^N \frac{\kappa}{2a^2}(x_i-x_{i+1})^2+|V_{\rm 1D}|\sin^2\left(\frac{\pi}{L}x_i\right),
\end{equation}
where $a$ is the horizontal spacing of nodes, $x_{N+1}=x_1$ and $\lambda_{k+1}=4\sqrt{\kappa/a^2}\sin^2(k\pi/N)$\cite{Braun2004}. It is well known that equilibrium averages may be obtained by ergodicity from stochastically integrating the overdamped Langevin equation\cite{Zwanzig}
\begin{equation}
\gamma\dot{x}_i = -\frac{\partial U(\bf x)}{\partial x_i} + \sqrt{2\gamma k_{\rm B}T}\eta_i(t)\\
\end{equation}
where the $\{\eta_i(t)\}_{i=1}^N$ are Gaussian random variables of zero mean and variance $\langle\eta_i(t)\eta_j(t')\rangle=\delta_{ij}\delta(t-t')$. Let $a,L$=1 and choose $\gamma$ for numerical stability. To show agreement with traditional transition state theory, I set the line tension $\kappa$=300meV to be much larger than the particle barrier $|V_{\rm 1D}|$=15meV; when $\kappa$ and $|V_{\rm 1D}|$ are comparable, the discrete structure produces a significant kink migration barrier whose effects are reported in detail elsewhere\cite{swinburne2013}.
Whilst the choice of energy units makes these numerical values appropriate for a dislocation line the phenomenology the model exhibits is general and widely reported\cite{Braun1998}. In particular, the exponential prefactor becomes inversely length dependent due to the lack of any Goldstone mode\cite{KMnote}.

Using a high quality random number generator\cite{dSFMT} to produce trajectories of $\sim10^{11}$ timesteps, the average value of a function $f({\bf x})$ was recorded for a value of $\bar x\in[0,L]$ to produce a Monte-Carlo evaluation of $\langle f({\bf x});\bar x\rangle$. To evaluate the free energy $F_{\rm U}(\bar x)$ a histogram of center of mass values $\bar x\in[0,L]$ was populated to produce $Z_\lambda\langle\exp(-\beta U);{\bar x}\rangle=\exp(-\beta F_{\rm U}(\bar x))$.\\
The results of these simulations are displayed in Figure (\ref{sim_plot}), showing that the diffusivity is indeed bounded by (\ref{highbound}) and (\ref{lowbound}). The free energy upper bound can be seen to provide a reasonable and qualitatively accurate approximation to the diffusivity at intermediate temperatures and, importantly, gives the correct activation energy at low temperature. The high temperature expansion (\ref{hightemp}) also becomes increasingly accurate once the thermal energy exceeds the particle barrier such that the convergence criterion $\beta|V_{\rm 1D}|< 1$ is satisfied.
\section{Non-Linear Response}\label{sec:nlr}
\begin{figure}[!t]
\includegraphics[width=0.48\textwidth]{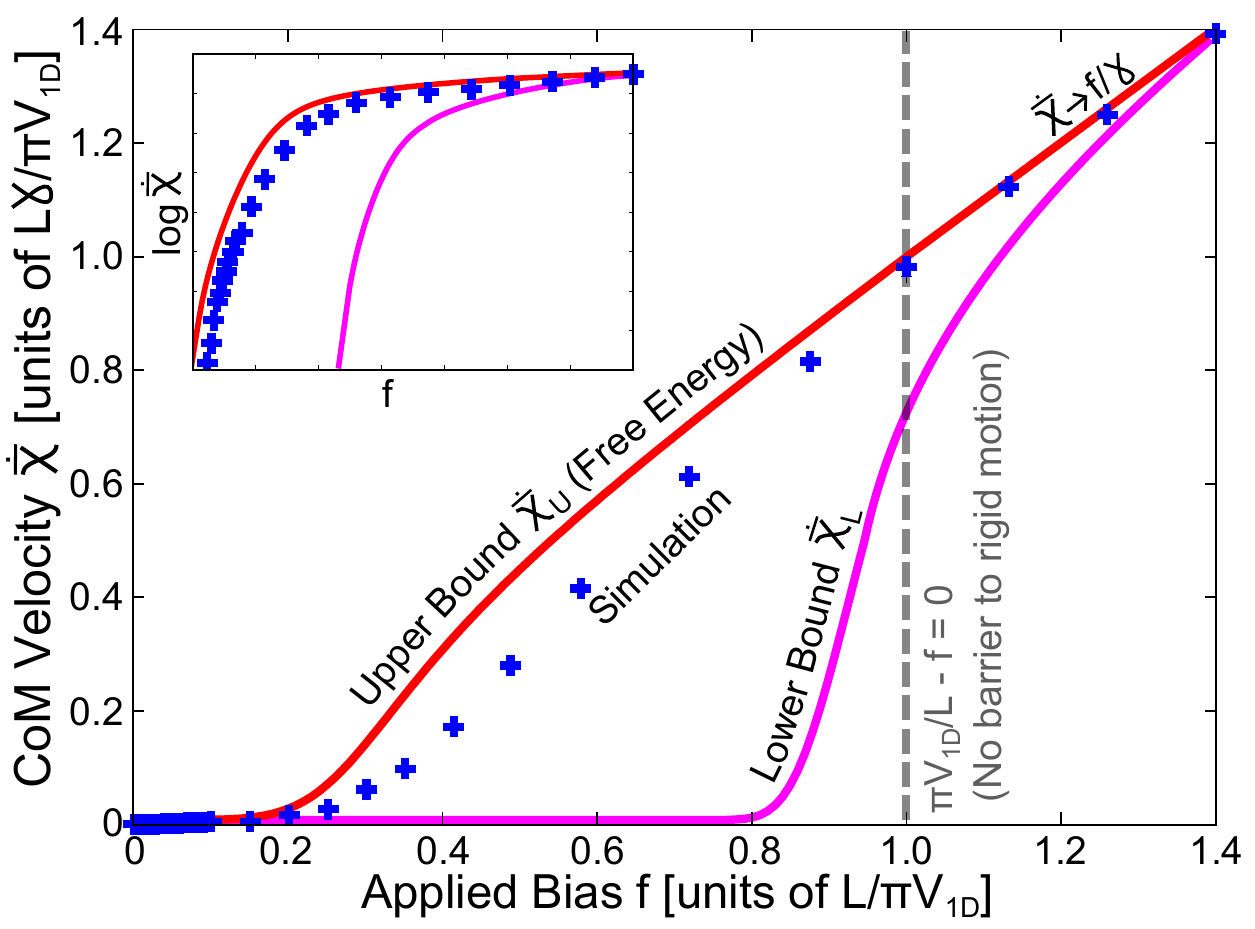}
\caption{(Color online) Non-linear response the same 40 particle Sine-Gordon chain as above at low temperature $k_{\rm B}T=6$meV. Inset: log plot showing the low bias response. Whilst the lower bound of (\ref{mu}), $\dot{\bar \chi}_{\rm L}$, only agrees at the highest bias, the `free energy' upper bound $\dot{\bar \chi}_{\rm U}$, is seen to show good agreement. The applied bias is expressed in proportion to the maximum gradient of the sinusoidal substrate potential, $\pi V_{\rm 1D}/L$; when $f>1$ in these units the biased on site potential has no stationary points meaning a drift is expected even at eero temperature. As the bias increases further any affect of the on site potential disappears and one recovers the free drift law $\dot{\bar\chi}$=$f/\gamma$. \label{NLRR}}
\end{figure}
To end, a DC bias $f$ is applied to the FK chain, such that the one dimensional on-site potential becomes $V_{\rm 1D}(x)-fx$. The effect of this bias is to break the symmetry of the system, meaning that the center of mass will drift with a velocity $\dot{\bar\chi}$. In the absence of any on site potential, it is simple to show that the free drift velocity is $f/\gamma$. Stratonovich\cite{kusnev,stratonovich1967} found the response of an overdamped point particle to such a bias to be 
\begin{equation}
{\dot x}_{\rm 1D}(f)  = 
\frac{L\left(1-e^{-\beta fL}\right)/{\beta \gamma}}{ \oint e^{-\beta(V_{\rm 1D}({x})-f{x})} \int_{x}^{{x}+ L}e^{\beta (V_{\rm 1D}({y})-f{y})} {\rm d}{y} {\rm d}{x} }.\label{1dnlr}
\end{equation}
The effective one dimensional migration potentials $F_{\rm L,U}(\bar x)$ implied by the diffusivity bounds suggest bounds $\dot{\bar \chi}_{\rm L,U}(f)$ on the non-linear response, through analogy to the Stratonovich result (\ref{1dnlr})
\begin{equation}
\dot{\bar \chi}_{\rm L,U}(f)=
\frac{L\left(1-e^{-\beta NfL}\right)/{\beta N\gamma}}{ \oint e^{-\beta(F_{\rm U}({\bar x})-Nf\bar{x})} \int_{\bar x}^{{\bar x}+ L}e^{\beta (F_{\rm L,U}({\bar y})-Nf\bar{y})} {\rm d}{\bar y} {\rm d}{\bar x} }, \label{mu}
\end{equation}
These bounds have been compared to stochastic simulation as before; typical results are displayed in Figure (\ref{NLRR}). At low temperaure the true result is much closer to the `free energy' upper bound, which again agrees with the transition state theory approximation. At a given tempearature, the properties of $\dot{\bar \chi}_{\rm U}$ are identical to the point particle result, which is well documented\cite{Risken}. This informative approximation to the low temperature non-linear response can be calculated at zero temperature in regimes where transition state theory is expected to apply, as the free energy landscape (\ref{fsde}) can be calculated from a constrained static minimisation\cite{Proville2012}.
\section{Conclusions and Outlook}\label{sec:dis}
The main result of this paper is that for the simple and widely employed model studied, the Helmholtz free energy landscape only gives a lower bound for any migration barrier to bulk motion. This result was obtained through diffusivly scaling the adjoint Fokker-Planck equation to isolate the long time limit and confirmed through extensive numerical simulation. An analagous relationship was also seen to hold for the non-linear response. Recalling that the free energy is an entropic maximum, it is not altogether surprising that the free energy pathway provides an upper bound on the diffusive transport; due to the simplicity and generality of (\ref{pot}), these results will hold for a wide range of physical systems. 
In future work, it would be interesting, using the approach developed here, to quantify the affect of both intertia and general particle interaction on many-body, non-linear, stochastic transport.
\section{ACKNOWLEDGMENTS}
I would like to thank the referees for helpful comments and S L Dudarev and A P Sutton for stimulating discussions and critical reviews of an earlier manuscript. I was supported through a studentship in the Centre for Doctoral Training on Theory and Simulation of Materials at Imperial College London funded by EPSRC under Grant No. EP/G036888/1. This work, partially supported by the European Communities under the contract Association between EURATOM and CCFE, was carried out within the framework of the European Fusion Development Agreement. The views and opinions expressed herein do not necessarily reflect those of the European Commission. This work was also partly funded by the RCUK Energy Programme under Grant No. EP/I501045.

\appendix
\section{Proof of (\ref{CS2})}\label{sec:app}
As $\rho_\infty$ and the test functions $f,g$ are periodic and bounded in $\bar x$, we may always expand $\sqrt{\rho_\infty}f$ (or $\sqrt{\rho_\infty}g$) as
\begin{equation}
\sqrt{\rho_\infty}f = \sum_{n=0}^{n=\infty}\tilde{f}_n(\{a_k\})\cos(\frac{2\pi n}{L}{\bar x})+\tilde{f}_{n+N}(\{a_k\})\sin(\frac{2\pi n}{L}{\bar x})\label{func_exp},
\end{equation}
where we have suppressed any $\bar\chi$ or $t$ dependence as they may be considered constant in the following.
The normalisation condition (\ref{opernorm}) may now be writen as
\begin{equation}
\frac{L}{2}\sum_{n=0}^{n=\infty}\left(\int_{\{a_k\}}\tilde{f}_n^2(\{a_k\})+\tilde{f}_{n+N}^2(\{a_k\})\right)<\infty,
\end{equation}
implying that the real $\tilde{f}_n(\{a_k\})$ must be square integrable functions, i.e. that
\begin{equation}
\int_{\{a_k\}}\tilde{f}_n^2(\{a_k\})<\infty.
\end{equation}
This means the functions satisfy a Cauchy-Schwartz inequality of the form
\begin{equation}
\left(\int_{\{a_k\}}\tilde{f}_n\tilde{f}_m\right)^2\leq\left(\int_{\{a_k\}}\tilde{f}_n^2\right)\left(\int_{\{a_k\}}\tilde{f}_m^2\right),\label{CS3}
\end{equation}
where the arguments of the functions have been omitted for brevity. For each value of $\bar x\in[0,L]$ the trigonometric functions in (\ref{func_exp}) may be considered coefficients in a linear sum of square integrable functions. As any linear combination of square integrable functions is also a square integrable function, any two linear combinations will also satisfy a Cauchy-Schwartz inequality of the form (\ref{CS3}). Taking $\sqrt{\rho_\infty}f$ and $\sqrt{\rho_\infty}g$ for these two linear combinations gives the desired proof of the pointwise inequality (\ref{CS2}). Note that (\ref{CS2}) is not derived explicitly from (\ref{CS1}).

\begin{thebibliography}{42}
\bibitem [{PER()}]{PERnote}%
  \BibitemOpen
  \href@noop {} {}\bibinfo {note} {To see the $\bar x$-periodicity of any
  solution $\Phi({\bar\chi},{\bar x},\{a_{p>0}\})$, note that $\hat{\rm L}_{\rm
  aFP}$ acts on $({\bar x},\{a_k\})$ and is invariant under a shift in $\bar x$
  by $nL$, where $n$ is an integer. For the solution $\Phi({\bar
  x}+nL,\{a_{p>0}\})$ to be unique we thus require $\Phi({\bar\chi},{\bar
  x}+nL,\{a_{p>0}\})$=$\Phi({\bar\chi},{\bar x},\{a_{p>0}\})$, giving the
  desired result.}\BibitemShut {Stop}%
\bibitem [{POT()}]{POTnote}%
  \BibitemOpen
  \href@noop {} {}\bibinfo {note} {For the analagy to be complete we associate
  $Z$=$Z_\lambda\int_{\bar x}\langle\exp(-\beta V);{\bar x}\rangle$ for the
  chain with $Z$=$\int_{x}\exp(-\beta V_{\rm 1D}(x))$ for the point
  particle.}\BibitemShut {Stop}%
\bibitem [{CUR()}]{CURnote}%
  \BibitemOpen
  \href@noop {} {}\bibinfo {note} {We emphasise that the $N$-1 positive
  curvatures will not in general be diagonal in the $\{a_k\}$
  coordinates.}\BibitemShut {Stop}%
\bibitem [{CON()}]{CONVnote}%
  \BibitemOpen
  \href@noop {} {}\bibinfo {note} {As the energy of the system can at most
  increase linearly with $N$, the $n^{th}$ term in the expansion
  $\ln\langle\exp(\pm\beta V);{\bar x}\rangle$ is of order $N(\beta|V_{\rm
  1D}|)^n$, giving the convergence criteria $\beta|V_{\rm 1D}|<1$.}\BibitemShut
  {Stop}%
\end{thebibliography}
\end{document}